\documentclass{article}
\usepackage{spconf,amsmath,graphicx}
\usepackage{array} 
\usepackage[margin=1in]{geometry} 
\usepackage{abstract} 
\usepackage{setspace} 
\usepackage{amssymb}
\usepackage{subcaption}
\usepackage{caption}
\usepackage{url}
\usepackage{hyperref}
\usepackage{booktabs}
\title{PRODUCTIVITY OF SHORT TERM ASSETS AS A SIGNAL OF FUTURE STOCK PERFORMANCE}

\name{}
\address{}

\begin{document}
\ninept

\twocolumn[
\begin{@twocolumnfalse}
\maketitle
\vspace{-6em} 

\begin{center}
    \textit{December 16, 2024} 

    \vspace{2em} 
    \begin{tabular}{>{\centering\arraybackslash}p{6cm} >{\centering\arraybackslash}p{6cm}} 
        \textbf{Veer Vohra} & \textbf{Devyani Vij} \\
        Undergraduate, Columbia IEOR & Undergraduate, Columbia IEOR \\
        New York, NY 10027 & New York, NY 10027 \\
        \texttt{vv2344@columbia.edu} & \texttt{dv2485@columbia.edu} \\
    \end{tabular}
    
    \vspace{1em} 
    
    \begin{tabular}{>{\centering\arraybackslash}p{6cm} >{\centering\arraybackslash}p{6cm}} 
        \textbf{Arman Ozcan} & \textbf{Jehil Mehta} \\
        Undergraduate, Columbia CS & Graduate, Columbia CS \\
        New York, NY 10027 & New York, NY 10027 \\
        \texttt{ao2794@columbia.edu} & \texttt{jjm2266@columbia.edu} \\
    \end{tabular}
\end{center}

\vspace{1em}
\begin{abstract}
This paper investigates cash productivity as a signal for future stock performance, building on the cash-return framework of Faulkender and Wang (2006). Using financial and market data from WRDS, we calculate cash returns as a proxy for operational efficiency and evaluate a long-only strategy applied to Nasdaq-listed non-financial firms. Results show limited predictive power across the broader Nasdaq universe but strong performance in a handpicked portfolio, which achieves significant positive alpha after controlling for the Fama-French three factors. These findings underscore the importance of refined universe selection. While promising, the strategy requires further validation, including the incorporation of transaction costs and performance testing across economic cycles. Our results suggest that cash productivity, when combined with other complementary signals and careful universe selection, can be a valuable tool for generating excess returns.
\end{abstract}
\vspace{1em}
\end{@twocolumnfalse}
]

\section{Introduction}
\label{sec:intro}

Over the past two decades, corporate cash holdings have increased to unprecedented levels, reflecting a major shift in how firms approach financial management. For instance, data shows that cash reserves held by U.S. non-financial companies grew by \textit{117\%} between 2007 and 2014, climbing to nearly \$2 trillion by year-end. This trend highlights how companies are stockpiling cash to protect against economic uncertainty, take advantage of strategic opportunities, or adapt to low-yield environments. Although holding cash provides a buffer during periods of volatility and ensures flexibility for future investments, it also raises concerns about inefficiency and underutilized capital. 
\vspace{0.5\baselineskip}

\begin{figure}[h!]
    \centering
    \includegraphics[width=0.48\textwidth]{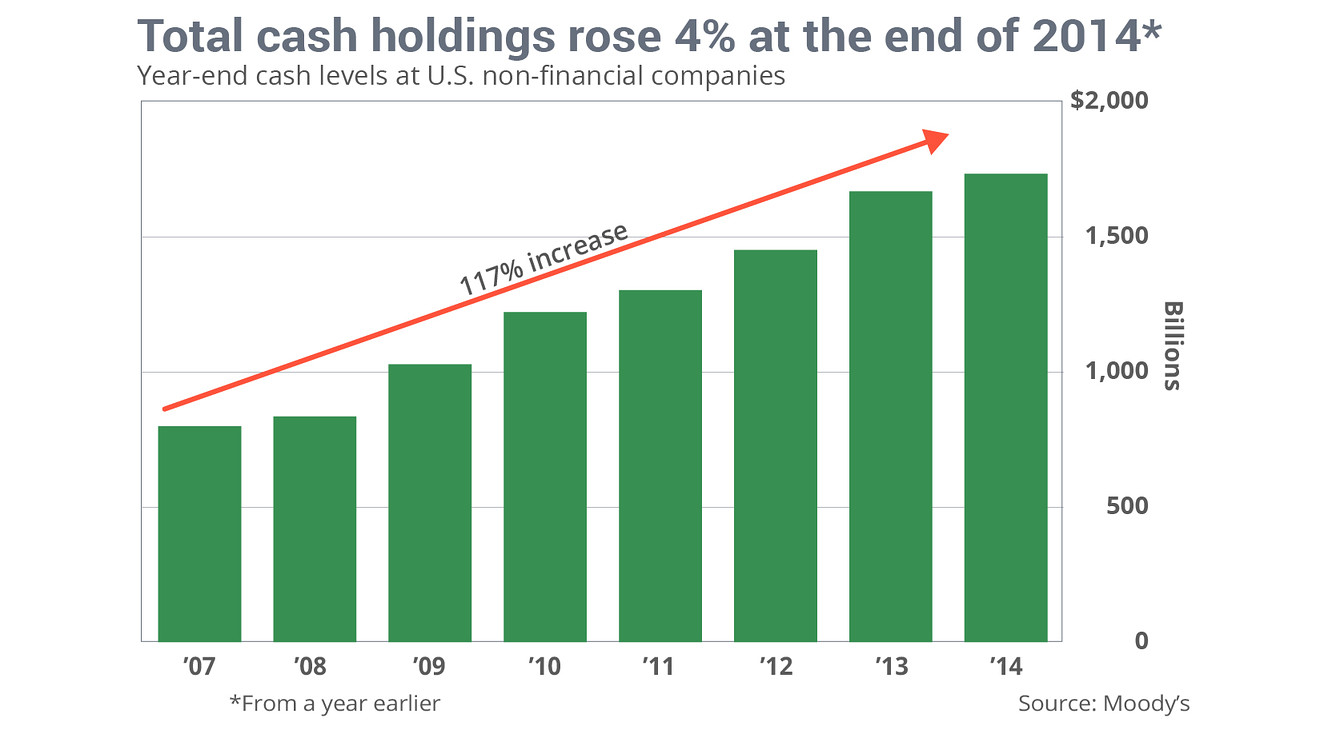}
    \caption{Year-end cash levels at U.S. non-financial companies, showing a 117\% increase from 2007 to 2014. Source: Moody's.}
    \label{fig:cash_holdings}
\end{figure}

The paper \textit{Cash-Hedged Stock Returns} \cite{ross2023} introduces a novel framework for understanding the role of corporate cash holdings in stock returns. By decomposing stock returns into cash and non-cash components, the study isolates the operational efficiency of firms from their implicit cash positions. This approach reveals that corporate cash holdings distort beta estimates and contribute to a higher covariance structure across stocks due to the correlated nature of cash returns. Importantly, the authors demonstrate that common asset pricing factors, such as size, value, and momentum, often include significant implicit cash positions, biasing factor returns and their performance.
\vspace{0.5\baselineskip}

Motivated by this framework, our paper seeks to replicate and validate the robustness of their findings on cash-hedged strategies. However, through an iterative process of refining the methodology, we modify the study by examining the role of cash-hedged returns as a signal for productivity. Specifically, we demonstrate that firms’ implicit cash returns not only hedge against volatility but also reflect underlying operational efficiency and productivity. By uncovering this link, positioning cash returns as a meaningful factor in evaluating firm performance and building productivity-driven portfolios.

\subsection{Initial Approach: Cash-Hedged Returns as a Signal}
Our starting point was to use \textit{cash-hedged returns} as a signal of company performance. Using the formulas outlined in the paper, we implemented calculations for cash-hedged returns (\( e_{it} \)) by:
\begin{itemize}
    \item Decomposing stock returns (\( r_{it} \)) into cash (\( b_{it} \)) and non-cash (\( e_{it} \)) components.
    \item Utilizing lagged data for cash weights (\( w_{it} \)) to ensure point-in-time accuracy and avoid look-ahead bias.
    \item Accounting for cash returns using simplified models based on risk-free rates and firm characteristics.
\end{itemize}
While the logic of focusing on the isolated performance of non-cash assets was sound, our empirical results revealed weaknesses. The cash-hedged returns lacked predictive power as a standalone signal for company performance. Despite their theoretical alignment with the paper’s framework, the implementation struggled to identify firms likely to deliver strong future returns.

\subsection{Re-Evaluating the Approach}
Recognizing these challenges, we revisited our methodology by closely inspecting the code and computations. During this process, we identified opportunities for refinement, particularly in how we derived and applied the cash-related components of stock returns. This led us to pivot toward a more fundamental approach: using \textit{cash returns} (\( b_{it} \)) directly as a signal.

\subsection{Shift to Cash Returns as a Signal}
Our revised methodology focused on cash returns as a proxy for company performance. The rationale was that firms utilizing their cash efficiently---whether for reinvestments, operational scaling, or strategic financial decisions---would likely outperform peers with less effective cash management. This logic aligns with practices employed by leading fundamental analysts, who view cash flow and resource allocation as core indicators of a firm’s health and future growth potential.

In practice, our code implemented this shift by:
\begin{enumerate}
    \item \textbf{Calculating Cash Returns (\( b_{it} \)):}
    \begin{itemize}
        \item Using lagged data for cash positions and total assets to derive cash weights (\( w_{it} \)).
        \item Modeling cash returns based on a simplified framework influenced by the firm-specific characteristics outlined in the paper.
    \end{itemize}
    \item \textbf{Integrating Cash Returns into Portfolio Construction:}
    \begin{itemize}
        \item Ranking firms based on their cash returns to construct signal-driven portfolios.
        \item Weighting the portfolio based on marginal cash productivity.
    \end{itemize}
\end{enumerate}

\section{Data}
\label{sec:data}

All of the data used in this paper is sourced through the Wharton Research Data Services (WRDS). This platform is a resource for datasets across the finance, economics, healthcare, marketing, and more industries for empirical research and analysis. This section outlines the data sourcing, selection, and preparation processes, emphasizing the measures taken to ensure reliability.

\subsection{Sourcing Historical Data}

In this study, we used public data from three vendors who publish their datasets on WRDS: Fama French Portfolios \& Factors, the Center for Research in Security Prices LLC (CRSP), and Computstat - Capital IQ. 
\begin{itemize}
    \item We extracted the daily risk-free rate and 3 Fama French Factors -market return, size premium, and value premium- from Fama French's daily factors table.
    \item The daily price and return of the S\&P500 index, the monthly return values of Nasdaq composite index, and the list of open trading days were pulled from CRSP.
    \item Company data was from both CRSP and Computstat; the former lent to the companies' daily stock prices, names, and unique identifiers (GVKEY and PERMNO), and the later stored financial data. This financial data was a compilation of the companies' 10Q releases over time and contained information about the reporting date, total assets, cash holding, total debt, earnings, research and development expenses, dividends paid, and interest expenses.
\end{itemize}
All data was collected starting from January 1st, 2009.

\subsection{Company Selection}
We consider two distinct stock universes for our analysis:
\begin{itemize}
    \item \textbf{Handpicked Portfolio:} A subset of random companies (e.g., Alphabet, Amazon, Apple, Microsoft, Nvidia, Meta Platforms, Tesla, Netflix, Walmart, and Pfizer). This portfolio allows for controlled testing of the signal's efficacy within a small group of firms - this was initially used just to test for accuracy and that our algorithms worked but proved to be a valuable portfolio to compare against.
    \item \textbf{NASDAQ Universe:} The broader NASDAQ-listed non-financial firms. Financial firms were excluded due to their differing balance sheet structures, using SIC codes outside the range of 6000--6799. Companies were filtered to include only those with data available from January 2000 to December 2023.
\end{itemize}

For each universe, we obtained the corresponding PERMNO (CRSP identifier) and GVKEY (Compustat identifier) using the CRSP-COMPUSTAT link table.

\subsection{Data Processing and Preparation}

After sourcing the raw data, we processed and transformed it into a consistent and usable format for our analysis. This section outlines the key steps taken:

\begin{enumerate}
    \item\textbf{Standardizing Returns to Monthly Frequency}

    The dataset contained price and return data reported at varying frequencies, specifically daily and monthly, which required standardization. When daily prices were not available, daily returns were compounded each month from between the last trading days of each month to give us accurate, aligned, monthly returns.
    
    For price data, we sampled the adjusted close price at the last trading day of each month to align all data. Monthly returns could then be easily calculated as a percentage change. This ensured that both returns and prices for all securities were consistently measured at a monthly level, enabling accurate comparisons across assets.

    \item \textbf{Handling Quarterly Financial Data: Point-in-Time Adjustments}  

    Financial data, such as cash holdings, is reported in companies' quarterly 10-Q filings. Due to differing fiscal calendars, these filings may be released at irregular intervals. This works to our advantage because even though data is released quarterly per company, we have access to new information each month. Simultaneously, this is a disadvantage because we only get new information about a company 4 times a year.
    \vspace{0.5\baselineskip}
    
    To ensure Point-in-Time (PiT) accuracy, we applied a \( t+1 \) trading day lag to the reporting date of financial data (trade today on yesterday's information). We then mapped each adjusted reporting date to the corresponding month-end trading day using a predefined month-end trading day calendar. To ensure comprehensive month-end coverage, we generated a full range of month-end dates and merged it with the mapped financial data. Missing month-end data was forward-filled to propagate the most recent financial information until new data became available.
    \vspace{0.5\baselineskip}

    This process ensured that the dataset reflected only the information available to investors at the time, maintaining PiT accuracy and avoiding look-ahead bias in our cash-based trading strategy.

    \item \textbf{Adjusting Prices for Corporate Actions}  

    To ensure accuracy in price-based calculations, we accounted for corporate actions, including stock splits and dividends. Adjustments were applied to both stock prices and shares outstanding using the cumulative adjustment factors provided in the dataset. Specifically, the adjusted stock price was calculated as:
    \begin{equation}
        \text{Adj. Price} = \frac{\text{Raw Price}}{\text{cfacpr}},
        \label{eq:adjusted_price}
    \end{equation}
    
    where the cumulative adjustment factor (cfacpr) accounts for stock splits and dividend distributions. Similarly, shares outstanding were adjusted by multiplying shares outstanding by the cumulative adjustment factor for shares (cfacshr). Using these adjustments, market capitalization was recalculated as:
    
    \begin{equation}
        \text{Market Cap} = \text{Adj. Price} \times \text{Adj. Shares Outstanding}.
        \label{eq:market_cap}
    \end{equation}
    
    To align price and market cap data with month-end trading dates, we merged the adjusted data with a predefined month-end trading day calendar. Additionally, percentage stock returns were computed as the daily percentage change in adjusted stock prices. These adjustments ensured that the price and market capitalization data remained consistent over time and accurately reflected company valuations after corporate actions.

    \item \textbf{Calculating Signal Variables}  

    Our implementation of the cash return strategy is based on Equation (9) in Faulkender and Wang (2006)\cite{faulkender2006}. This regression estimates the sensitivity of excess returns to changes in cash holdings and other financial variables. The model can be expressed as:

    \begin{align}
        r_{it} - R_t = \alpha + \gamma_1 \Delta \text{Cash}_{it} + \gamma_2 \Delta \text{Earnings}_{it} \notag \\ + \gamma_3 \Delta \text{Assets}_{it} + \ldots + \epsilon_{it}
        \label{eq:cash_strategy}
    \end{align}
    where \( r_{it} \) is the stock return for company \( i \) at time \( t \), \( R_t \) is the risk-free rate, and \( \Delta \text{Cash}_{it} \), \( \Delta \text{Earnings}_{it} \), and other terms capture changes in financial metrics.
    
    In our modified implementation, we scale each change by the lagged market capitalization \( M_{t-1} \) to normalize values relative to firm size. The regression is performed using the following variables:
    \[
    \begin{array}{ll}
    \gamma_1 = \frac{\Delta \text{Cash}_{it}}{M_{t-1}} & \gamma_2 = \frac{\Delta \text{Earnings}_{it}}{M_{t-1}} \\
    \gamma_3 = \frac{\Delta (\text{Assets}_{it} - \text{Cash}_{it})}{M_{t-1}} & \gamma_4 = \frac{\Delta \text{R\&D Expense}_{it}}{M_{t-1}} \\
    \gamma_5 = \frac{\Delta \text{Interest Expense}_{it}}{M_{t-1}} & \gamma_6 = \frac{\Delta \text{Dividends Paid}_{it}}{M_{t-1}}
    \end{array}
    \tag{6}
    \]
    
    \noindent  We also include nonlinear terms to capture interactions between cash holdings and leverage
    
    \[
    \begin{array}{ll}
    \gamma_7 = \frac{\text{Cash Holdings}_{t-1}}{M_{t-1}} & \gamma_8 = \text{Leverage}_{it} \\
    \gamma_9 = \frac{\Delta (\text{Total Debt}_{it} + M_{t-1})}{\text{Lagged Total Debt}_{it} + M_{t-1}} & \gamma_{10} = \frac{M_{t-1} \cdot \Delta \text{Cash}_{it}}{M_{t}^2} \\
    \gamma_{11} = \frac{\text{Leverage}_{it} \cdot \Delta \text{Cash}_{it}}{M_{t}} & 
    \end{array}
    \tag{7}
    \]
    
    The regression is estimated using Ordinary Least Squares (OLS), and the coefficients are used to compute the marginal value of cash. Specifically, the marginal cash value is given by:
    \begin{align}
        \text{Marginal Cash Value} &= \alpha 
        + \gamma_7 \frac{\text{Cash Holdings}_{t-1}}{M_{t-1}} \notag \\
        &\quad + \gamma_8 \text{Leverage}_{it}.
    \end{align}
    
    The average cash value is obtained by multiplying the marginal cash value by the current cash holdings. Finally, the monthly cash return \( b_{it} \) is computed as the percentage change in the average cash value:
    \begin{equation}
        b_{it} = \frac{\text{Average Cash Value}_{t} - \text{Average Cash Value}_{t-1}}{\text{Average Cash Value}_{t-1}}
    \end{equation}
    
    This approach allows us to quantify the contribution of cash holdings and other financial metrics to a firm's excess returns while accounting for firm size and leverage effects. To mitigate the impact of extreme values, we winsorized the cash return distribution at the 1st and 99th percentiles. Winsorization reduces the influence of outliers while preserving the structure of the data, ensuring robust signal calculations.

\end{enumerate}
Through these steps, we transformed raw financials, price, and return data into a standardized, Point-in-Time-adjusted, and clean dataset. The processed data enabled the implementation of our trading strategy and ensured consistency across all companies and time periods.

\section{Portfolio Implementation}
\label{sec:portfolio-implementation}
The backtesting framework ensures that all decisions are made based on data available at the time of investment (Point-in-Time accuracy). The portfolio weights are derived directly from the predictive signal, and the strategy is evaluated against widely recognized benchmarks to assess its relative performance. Optimization of the lookback period further refines the signal's implementation, demonstrating its robustness under realistic backtesting conditions.

\subsection{Backtesting Framework}

To evaluate the predictive power of the cash return signal \( b_{it} \), we implemented a backtesting framework that simulates a monthly rebalancing strategy. The backtesting process involves the following steps:

\begin{enumerate}
    \item \textbf{Signal Construction:} For each firm, the cash return signal \( b_{it} \) is calculated based on changes in cash holdings and other financial metrics, as previously described. 
    \item \textbf{Lookback Period:} For each month \( t \), a rolling lookback window of \( L \) months is used to compute the average signal performance for each firm. Firms with a positive average signal (\( b_{it} > 0 \)) are included in the portfolio.
    \item \textbf{Weight Assignment:} Portfolio weights are assigned based on the magnitude of the signal performance during the lookback period. The weights are normalized to ensure they sum to one:
    \begin{equation}
        w_{i,t} = \frac{\text{Signal}_{i,t}}{\sum_{j \in P_t} \text{Signal}_{j,t}},
    \end{equation}
    where \( w_{i,t} \) is the weight of firm \( i \) in month \( t \), and \( P_t \) represents the set of firms with a positive signal at time \( t \).
    \item \textbf{Portfolio Return:} The portfolio's return for month \( t \) is computed as the weighted sum of individual stock returns:
    \begin{equation}
        R_{p,t} = \sum_{i \in P_t} w_{i,t} \cdot R_{i,t+1},
    \end{equation}
    where \( R_{i,t+1} \) is the return of firm \( i \) in the following month.
\end{enumerate}

\subsection{Lookback Period Optimization}

The lookback period \( L \), which determines the length of historical data used to compute the signal, was optimized to maximize the Sharpe ratio of the portfolio during a training period (2010--2015). Using the Powell optimization method, the Sharpe ratio was defined as:
\begin{equation}
    \text{Sharpe Ratio} = \frac{\mathbb{E}[R_{p,t}]}{\text{std}(R_{p,t})},
\end{equation}
where \( \mathbb{E}[R_{p,t}] \) is the mean excess return and \( \text{std}(R_{p,t}) \) is its standard deviation. The optimal lookback period was optimized from 2010-2015 (our training data) for each portfolio. Optimized values ranged from 3 to 11.

\subsection{Testing and Performance Evaluation}

The optimized lookback period was applied to an out-of-sample testing period from January 2015 onward. The portfolio's cumulative returns were compared to the following benchmarks:
\begin{itemize}
    \item The S\&P 500 Index,
    \item The NASDAQ Composite Index,
    \item The risk-free rate (extracted from Fama-French data).
\end{itemize}

Cumulative returns were calculated as:
\begin{equation}
    R_{\text{cum},t} = \exp\left( \sum_{\tau=0}^t \log(1 + R_{p,\tau}) \right) - 1.
\end{equation}

The backtesting results include the monthly portfolio returns, cumulative returns, and comparisons to benchmarks, allowing for a comprehensive evaluation of the cash return signal's performance.

\section{Results}
\label{sec:results}

Our returns analysis demonstrates that while the strategy yields promising outcomes under a carefully curated portfolio, its performance suffers when applied to broader indices, such as the NASDAQ. The discrepancy highlights that the universe selection and narrowing process plays a critical role in achieving superior results, but also may indicate a larger issue with the signal.

\subsection{NASDAQ Portfolio Performance}
The results for the NASDAQ portfolio across multiple lookback month options reveal generally weak performance, as illustrated in the following graphs:

\begin{figure}[h!]
    \centering
    \begin{subfigure}[b]{0.45\textwidth}
        \includegraphics[width=\textwidth]{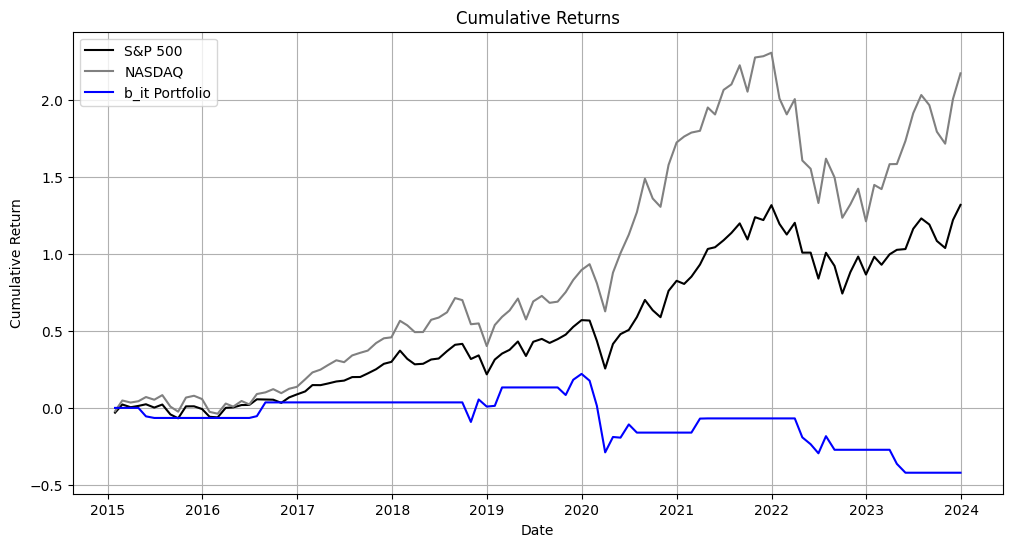}
        \caption{Cumulative Returns}
        \label{fig:nasdaq_cum_ret}
    \end{subfigure}
    \hfill
    \begin{subfigure}[b]{0.45\textwidth}
        \includegraphics[width=\textwidth]{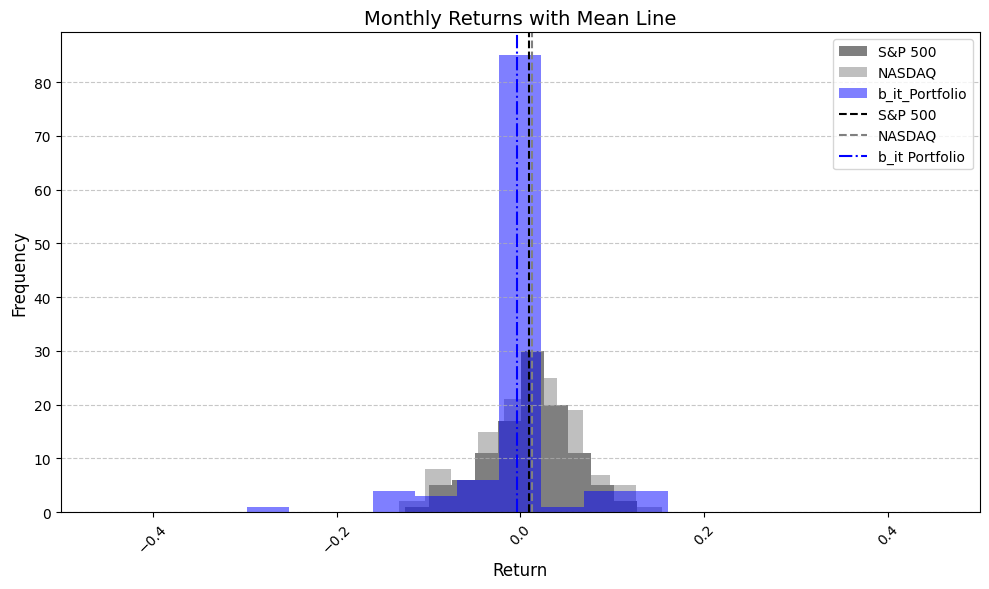}
        \caption{Monthly Return Distribution}
        \label{fig:nasdaq_monthly_hist}
    \end{subfigure}
    \vfill
    \begin{subfigure}[b]{0.45\textwidth}
        \includegraphics[width=\textwidth]{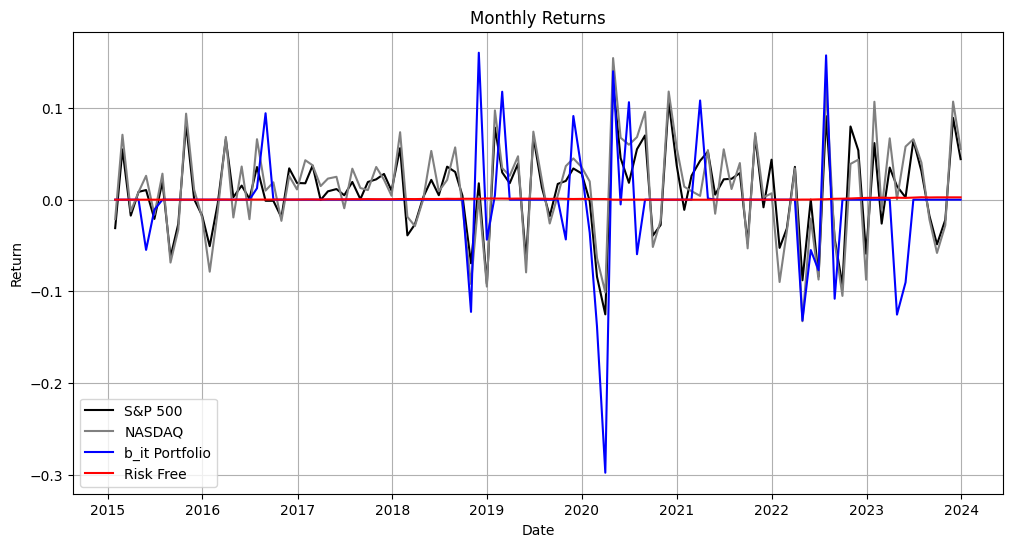}
        \caption{Monthly Return Trends}
        \label{fig:nasdaq_monthly_line}
    \end{subfigure}
    \caption{NASDAQ Portfolio Monthly and Cumulative Returns}
    \label{fig:nasdaq_performance}
\end{figure}

\textbf{Key Metrics:}
\begin{itemize}
    \item \textbf{Sharpe Ratio:} Relatively low, reflecting poor risk-adjusted returns.
    \item \textbf{Alpha:} Minimal or negative, confirming that the strategy struggles to generate excess returns in the broader NASDAQ universe.
\end{itemize}

Previously, we had optimized for the best lookback period over the entire timeframe, which manufactured better results as everything was in-sample with data leakages from using future values for this hyperparameter tuning. These results, while not as optimistic about using cash productivity as a signal, are a result of out-of-sample testing. 
\vspace{0.5\baselineskip}

This discrepancy between in-sample and out-of-sample testing could indicate two different conclusions. Optimistically, this could mean that even if the standalone measure of historical cash productivity is not enough, a well-developed machine learning model that predicts future cash productivity could yield better results, more similar to the in-sample ones. However, a larger future study would have to be conducted. Taking these results as they are, past cash productivity as it is constructed in this paper is not a strong enough standalone signal to select high return stocks in a wide array of companies such as the NASDAQ.

\subsection{Handpicked Portfolio Performance}
In contrast, the handpicked portfolio of companies exhibits \textbf{strong performance}, underscoring the value of a refined universe selection process:

\begin{figure}[h!]
    \centering
    \begin{subfigure}[b]{0.45\textwidth}
        \includegraphics[width=\textwidth]{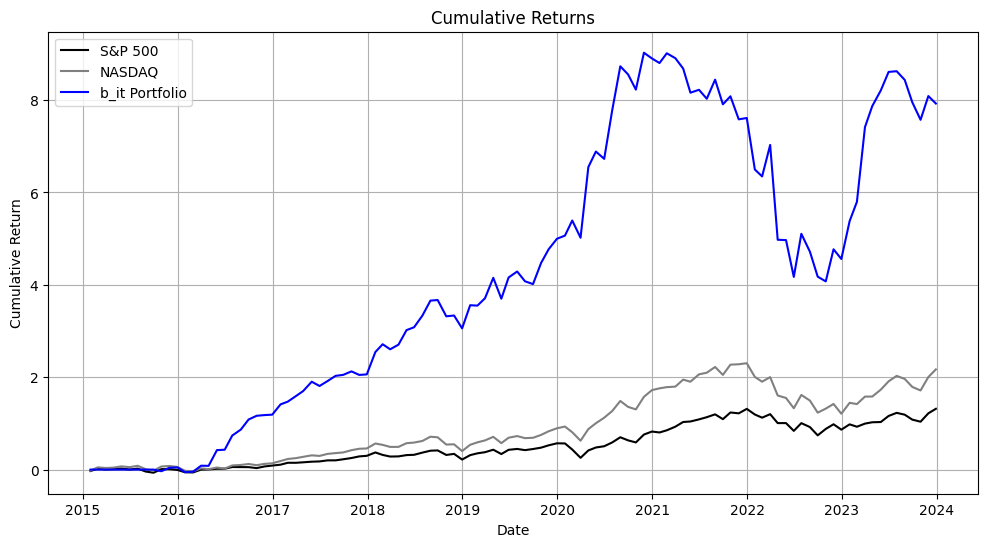}
        \caption{Cumulative Returns}
        \label{fig:handpicked_cum_ret}
    \end{subfigure}
    \hfill
    \begin{subfigure}[b]{0.45\textwidth}
        \includegraphics[width=\textwidth]{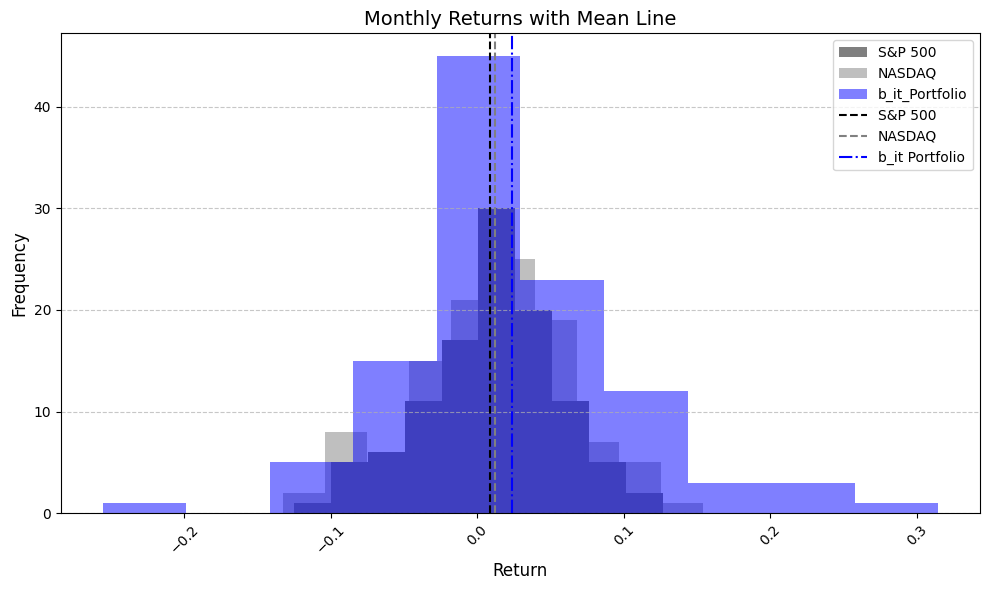}
        \caption{Monthly Return Distribution}
        \label{fig:handpicked_monthly_hist}
    \end{subfigure}
    \vfill
    \begin{subfigure}[b]{0.45\textwidth}
        \includegraphics[width=\textwidth]{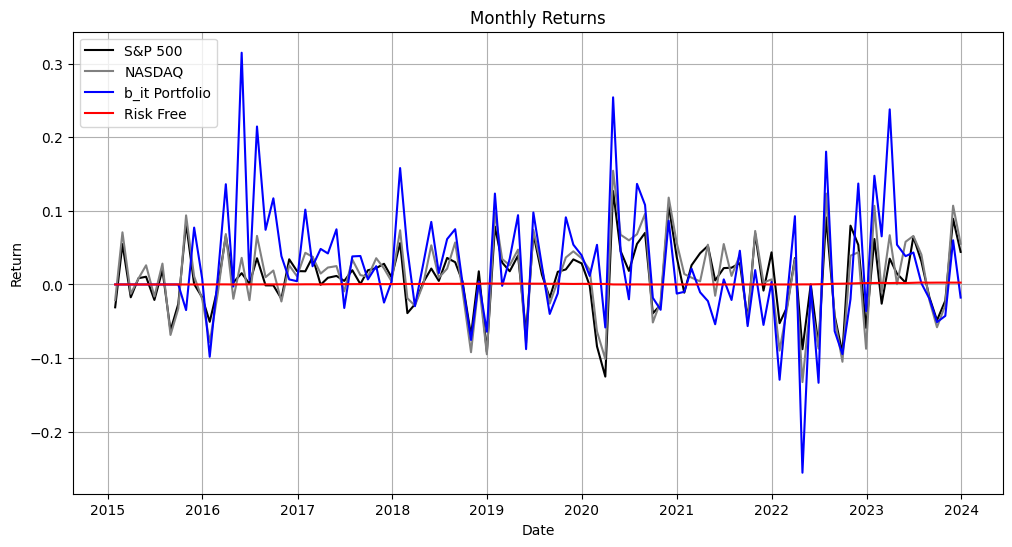}
        \caption{Monthly Return Trends}
        \label{fig:handpicked_monthly_line}
    \end{subfigure}
    \caption{Handpicked Portfolio Monthly and Cumulative Returns}
    \label{fig:handpicked_performance}
\end{figure}

\textbf{Key Metrics:}
\begin{itemize}
    \item \textbf{Sharpe Ratio:} High, indicating strong risk-adjusted performance.
    \item \textbf{Alpha:} Significant positive alpha, suggesting that the handpicked portfolio generates meaningful excess returns after controlling for market factors.
\end{itemize}

\begin{table}[h!]
    \centering
    \caption{\textbf{Portfolio Performance Metrics}}
    \label{tab:portfolio_performance}
    \resizebox{\columnwidth}{!}{%
    \begin{tabular}{lcccc}
        \toprule
        \textbf{Asset} & \textbf{Mean} & \textbf{Volatility} & \textbf{Sharpe} & \textbf{Alpha} \\ 
        \midrule
        \textbf{S\&P 500}                  & 0.1114 & 0.0252 & 0.6255 & 0.0008 \\
        \textbf{NASDAQ}                   & 0.1565 & 0.0357 & 0.7417 & 0.0029 \\
        \textbf{Fixed NASDAQ Portfolio}   & -0.1715 & 0.1802 & -0.4754 & -0.0134 \\
        \textbf{Handpicked Portfolio}     & 0.3216 & 0.0790 & 0.9952 & 0.0147 \\
        \bottomrule
    \end{tabular}%
    }
\end{table}

This portfolio is generating positive alpha,  which was calculated by regressing against the benchmark and the 3 fama french factors [features were standardized for enhanced comparability]. This was done to make sure that the calculated alpha was not a result of capturing other prominent factors and is indeed due to our isolated strategy. As shown in Table 2, alpha is statistically significant, as are the market and value factors. This lends to the need to test portfolios that use cash-returns as one of many signals. 

\begin{table}[h!]
    \centering
    \caption{OLS Regression Results}
    \label{tab:ols_results}
    \resizebox{\columnwidth}{!}{%
    \begin{tabular}{lcccccc}
        \toprule
        \textbf{Variable} & \textbf{Coefficient} & \textbf{Std. Err} & \textbf{t-Statistic} & \textbf{P-value} & \textbf{[0.025]} & \textbf{[0.975]} \\ 
        \midrule
        \textbf{const}            & 0.0165   & 0.006   & 2.635   & 0.010   & 0.004   & 0.029 \\
        \textbf{market\_factor}   & 1.6818   & 0.271   & 6.202   & 0.000   & 1.144   & 2.220 \\
        \textbf{size\_factor}     & -0.4822  & 0.445   & -1.083  & 0.282   & -1.366  & 0.401 \\
        \textbf{value\_factor}    & -1.0752  & 0.280   & -3.846  & 0.000   & -1.630  & -0.521 \\
        \textbf{momentum\_factor} & -0.2075  & 0.314   & -0.660  & 0.511   & -0.831  & 0.416 \\ 
        \midrule
        \multicolumn{7}{l}{\textbf{R-squared:} \hspace{0.5em} 0.400 \hfill \textbf{Adjusted R-squared:} \hspace{0.5em} 0.377} \\
        \bottomrule
    \end{tabular}%
    }
\end{table}

\section{Discussion}
\label{sec:discussion}

This section discusses the limitations, robustness checks, and potential extensions of our study, as well as broader implications of the findings. Our iterative approach allowed us to refine the implementation of a cash-based trading strategy while ensuring that the results align with the underlying logic presented in the introduction.

\subsection{Robustness and Limitations of the Implementation}

Several checks remain necessary to ensure the robustness of cash productivity as a factor. First, the current portfolio implementation does not account for transaction costs incurred during monthly rebalancing or the tax implications of frequently realizing gains. Incorporating realistic trading frictions, such as bid-ask spreads, commissions, and capital gains taxes, would provide a more accurate assessment of the strategy's net performance.
\vspace{0.5\baselineskip}

The portfolio is implemented as long-only, which assumes the investor takes positions only in firms deemed cash-productive. However, this approach could potentially benefit from shorting firms with cash-unproductive characteristics. Exploring this extension could improve the strategy's risk-adjusted returns, particularly during market downturns or periods of heightened dispersion in cash productivity across firms.
\vspace{0.5\baselineskip}

Another important limitation is the time-frame and economic cycle testing. The current backtest was conducted over a single window, and its robustness across different economic phases—such as expansions, recessions, or high-interest rate environments—remains untested. Changes in government regulation, corporate behavior, or macroeconomic factors (e.g., interest rates) over time could alter the incentive for firms to hold significant cash balances. A comprehensive robustness analysis across multiple sub-periods would provide stronger evidence for the factor's durability.

\subsection{Liquidity Constraints and Real-World Considerations}

Our backtesting assumes that the investor has unlimited capital and liquidity to allocate to new positions as soon as a signal indicates a buy opportunity. In reality, investors face capital constraints and may need inflows of funds to maintain a fully invested portfolio without liquidating other positions. This assumption simplifies the analysis but overlooks potential liquidity issues that arise when implementing such a strategy at scale.
\vspace{0.5\baselineskip}

Additionally, practical considerations such as position size limits or constraints on purchasing illiquid stocks must be addressed. Future research could explore these dynamics by incorporating position-sizing rules, liquidity-adjusted weights, and capital allocation constraints into the backtesting framework.

\subsection{Reliance on Regression and Data Quality}

The calculation of cash returns relies heavily on regression modeling, which serves as a proxy for productivity within a company. While our implementation follows the methodology outlined in Faulkender and Wang (2006)\cite{faulkender2006}, future work could explore alternative modeling approaches, such as machine learning regressors (e.g., Random Forest or XGBoost), to determine whether these methods yield superior results in terms of out-of-sample performance or cross-validation accuracy. These advanced techniques may capture non-linear relationships or interactions between variables that are not well-represented in linear regression models.
\vspace{0.5\baselineskip}

\textbf{Forward-Filling and Imputation Bias}
In this study, forward-filling was applied to propagate financial data between reporting periods to maintain continuity. While this approach ensures that only Point-in-Time (PIT) information is used, it may introduce bias over longer time intervals, particularly for firms with sporadic reporting schedules or extended gaps in data availability. This limitation could affect the reliability of our cash productivity signal. Future work could explore alternative imputation methods, such as linear interpolation or machine learning-based imputations (e.g., k-nearest neighbors or predictive modeling), to reduce reliance on forward-filling while maintaining PIT accuracy and avoiding future data leakage.
\vspace{0.5\baselineskip}

\textbf{Winsorization Justification}
To mitigate the influence of extreme values, we winsorized the cash return distribution at the 1st and 99th percentiles. While this step reduces the impact of outliers, it is important to justify this choice and assess its effect on the results. Future iterations of this study should test the sensitivity of results to different winsorization thresholds (e.g., 5th and 95th percentiles) to ensure robustness and validate that the choice of threshold does not disproportionately affect the outcomes.
\vspace{0.5\baselineskip}

\textbf{Impact of Missing Data}
The WRDS dataset used in this study contains frequent missing values for critical features such as dividends, interest expenses, and cash holdings. To address this issue, we imputed zeros for non-critical variables and removed rows with missing values for critical variables like cash holdings. While this approach ensures a cleaner dataset, it may introduce bias by excluding firms with incomplete data. A more rigorous analysis of the impact of missing data on our results is warranted. For instance, future research could compare results obtained using different imputation techniques or evaluate whether firms with missing data exhibit systematic differences from those included in the analysis.

\subsection{Broader Implications and Future Work}

Despite these limitations, our findings have important implications for investors and portfolio managers. If further investigation confirms that cash productivity is a robust factor, it could serve as a valuable addition to multi-factor portfolios or as a standalone strategy for specific types of investors. Future work should address several key areas:
\begin{enumerate}
    \item\textbf{Imputation Methods:} Explore advanced imputation techniques to handle missing data more effectively while maintaining PIT accuracy.
    \item\textbf{Sensitivity Analysis:} Conduct sensitivity analyses on winsorization thresholds to ensure robustness against outliers.
    \item\textbf{Sector-Specific Effects:} Investigate whether cash productivity signals vary across industries or sectors.
    \item\textbf{Economic Cycle Testing:} Evaluate the robustness of cash productivity as a factor across different economic phases (e.g., recessions, expansions).
    \item\textbf{Macroeconomic Interactions:} Examine how macroeconomic variables like interest rates or inflation interact with cash productivity signals.
    \item\textbf{Realistic Portfolio Constraints:} Incorporate transaction costs, liquidity constraints, and position-sizing rules into backtesting frameworks to better reflect real-world trading conditions.
\end{enumerate}

By addressing these considerations, future research can refine the strategy into a more practical and reliable investment tool while enhancing its theoretical foundation. This revision integrates your requested points seamlessly into the discussion section while maintaining clarity and rigor. It highlights limitations transparently and provides actionable suggestions for future research directions.

\section{Summary}
\label{sec:summary}

This paper explores a quantitative approach to fundamental analysis by examining the relationship between corporate cash productivity and stock performance. We explore the use of cash productivity as a signal for future stock performance, motivated by the increasing importance of corporate cash holdings in financial management. Building on the framework of Faulkender and Wang (2006)\cite{faulkender2006}, we construct a cash return signal (\( b_{it} \)) as a proxy for firms' operational efficiency and evaluate its ability to generate excess returns.
\vspace{0.5\baselineskip}

To ensure accuracy, we applied Point-in-Time adjustments, forward-filled non-critical missing data, and winsorized outliers at the 1st and 99th percentiles. A backtesting framework was implemented, where portfolios were rebalanced monthly based on firms' cash return signals. We optimized the lookback window for signal calculation and conducted both in-sample and out-of-sample testing to ensure robustness.
\vspace{0.5\baselineskip}

Our results demonstrate a clear discrepancy between two portfolio implementations:
\begin{enumerate}
    \item The \textbf{handpicked portfolio} of high-signal firms generated significant positive alpha and high Sharpe ratios, highlighting the value of refined universe selection.
    \item The broader \textbf{NASDAQ portfolio} exhibited weaker performance, suggesting that cash productivity alone is insufficient as a standalone signal across a wide array of companies.
\end{enumerate}
Regression analysis using the Fama-French three factors confirms that the observed alpha is not explained by size, value, or market effects. While these findings validate the potential of cash productivity as an investment factor, we acknowledge several limitations, including the need to incorporate transaction costs, test across economic cycles, and explore alternative modeling approaches.
\vspace{0.5\baselineskip}

Future work should focus on improving data quality, testing the signal's performance across sectors, and integrating it into multi-factor models. If further validated, cash productivity could serve as a valuable addition to quantitative investment strategies, particularly for identifying firms with efficient cash management practices.

\section{Acknowledgments}
\label{sec:acknowledgments}

This study would not have been possible without the expertise and guidance of Professor Naftali Cohen. Through carefully curated lecture topics that informed our every step of the implementation, to direct feedback on our progress and direction refinements during class updates and office hours, to sorting out the free student subscription to access all of the data utilized in this paper, Professor Cohen has been an incredible resource to us. 

\section{Code}
The code for this project can be accessed at the link \href{https://github.com/veervohra03/QR-Cash-Returns}{here}.

\end{document}